# Which Values Matter to Socially Assistive Robots in Elder Care Settings? Empirically Investigating Values That Should Be Embedded in SARs from a Multi-Stakeholder Perspective

✉ Vivienne Jia Zhong, Institute for Information Systems, School of Business, University of Applied Sciences and Arts Northwestern Switzerland, Basel-Stadt, Switzerland

viviennejia.zhong@fhnw.ch, ORCID: 0000-0003-0605-5291

Theresa Schmiedel, Institute of Business Information Technology, School of Management and Law, Zurich University of Applied Sciences, Winterthur, Switzerland

shhd@zhaw.ch

Abstract:
The integration of socially assistive robots (SARs) in elder care settings has the potential to address critical labor shortages while enhancing the quality of care. However, the design of SARs must align with the values of various stakeholders to ensure their acceptance and efficacy. This study empirically investigates the values that should be embedded in SARs from a multi-stakeholder perspective, including care receivers, caregivers, therapists, relatives, and other involved parties. Utilizing a combination of semi-structured interviews and focus groups, we identify a wide range of values related to safety, trust, care, privacy, and autonomy, and illustrate how stakeholders interpret these values in real-world care environments. Our findings reveal several value tensions and propose potential resolutions to these tensions. Additionally, the study highlights under-researched values such as calmness and collaboration, which are critical in fostering a supportive and efficient care environment. Our work contributes to the understanding of value-sensitive design of SARs and aids practitioners in developing SARs that align with human values, ultimately promoting socially responsible applications in elder care settings.

# 1. Introduction

The integration of socially assistive robots (SARs) into care settings [1] raises questions about how we can maintain the human elements of care while addressing the labor crisis the nursing sector faces [2]. SARs are designed to assist users through social interaction rather than just physical tasks [3]. A recent review reveals that the real-world deployment of SARs in elderly care centers accounts for the second largest share of the total SAR distribution [4]. Among these deployments in elderly care centers, SARs primarily provide entertainment (games, singing, etc.) [5], emotional companionship, and educational activities that stimulate cognitive abilities in residents [6]. They also facilitate communication with family via video calls, promote physical activity and enhance safety by monitoring residents [7] among other things. The use of SARs appears to have positive effects on the psychological and physiological state of the elder users [8–10]. They improve elderlies' mood [11] and promote behavioral engagement that tends to increase over time [12]. Beyond, interactions with SARs can also positively impact interactions with other people [13].

While technological advances are crucial to the enhanced capabilities and benefits of SARs, a technocentric view has faced strong criticism [14–16]. Focusing on the development of new algorithms, such as improving multi-party conversational capabilities of SARs with large language models [17] and optimizing robot navigation in populated environments [18], corresponds to technological determinism, which views technology as the primary driver of social progress. However, researchers have increasingly recognized that technology is socially constructed, meaning social factors, such as our needs and values, influence how technologies are designed, used, and understood [16, 19].

Indeed, research emphasizes the bidirectional relationship between society and technology [15, 20, 21], urging the inclusion of diverse social groups in the design of SAR applications [14, 15]. To ensure SARs are truly helpful, research is shifting from technical feasibility towards understanding user needs [22–26]. However, it still underemphasizes an aspect critical in meeting human needs [27], namely values. Values represent the desirable, what people subconsciously deem important [28, 29]. Human values lie at the heart of our interactions, influencing our behaviors and expectations [30, 31]. Neglecting human values in SAR design not only hinders user acceptance [32] but can also lead to unintended negative consequences such as loss of human contact, objectification of elderly people, deception, or emotional attachment [14, 33, 34]. For instance, while residents in a nursing home may appreciate the convenience of having a robot perform routine tasks, such as helping them turn on the light, this over-reliance on robots for simple tasks could negatively impact residents' mental stimulation and gradually erode the value of autonomy.

Investigating the values that are important to relevant stakeholders in care settings is therefore necessary to mitigate potential harms while realizing the promised benefits of SAR applications. However, while research on values provides a rich set of theoretical value frameworks for designing SARs within care settings [35, 36], there is little empirical evidence about how these values manifest in practice. At the same time, recent discussions highlight the need to consider contextual factors that shape the understanding and priority of values in real-world settings [37–39]. While the given value frameworks are valuable, empirical research is essential to uncover contextual values that may complement existing theoretical frameworks.

Existing research on values primarily takes an ethical perspective, focusing on the views of care-receivers and caregivers [40–42]. However, the concept of value itself is multifaceted, encompassing ethical values as well as values from specific groups, including organization and department [43, 44]. The introduction of SARs into elderly care centers affects a complex ecosystem of stakeholders, including not only care-receivers and caregivers but also relatives, healthcare institutions, robot manufacturers, policymakers, and others [45], each with their own unique set of values and priorities. This complexity requires a broader understanding of values [46] and an expansion of the scope of stakeholders that need to be considered in SAR applications [47].

The lack of research on values from an empirical view, including contextual and multi-stakeholder perspectives, poses significant challenges for designing human-robot interactions (HRI) that align with human values and thereby truly meet the needs of all stakeholders involved in a SAR application. To address these gaps, we present an empirical study that investigates relevant values for the design and development of a SAR for a nursing home setting. Our study yields a wide range of values, highlighting the complexities involved in creating value-sensitive HRI.

The remainder of this paper is structured as follows: Section 2 reviews existing literature on values in care settings and approaches to integrating values into HRI design. Section 3 outlines our research methodology. Section 4 presents the key findings from our empirical research. Finally, we discuss our research in Section 5 before concluding our work in Section 6.

## 2. Research Background

The concept of value is studied in various disciplines [28, 48, 49]. To accommodate the various dimensions of values, we adopted the definition proposed by Friedman et al. [29] that refers the concept of value to "what a person or group of people consider important in life". In this section, we illustrate how values in HRI design are approached from existing research.

## 2.1. Conceptual Approaches to Value-Sensitive HRI

A large number of frameworks discuss the ethical design of SARs in healthcare [50, 51]. One of them is the Care Centered Framework developed by van Wynsberghe [40], which draws attention to researchers [52, 53]. This framework is based on Value Sensitive Design (VSD) [29], an approach accounting for human values in technology design. The framework builds on care ethics and moral elements, as proposed by Tronto [54], and entails five components: the context, practice, key actors involved, robot's type, and the manifestation of care values [54]: attentiveness, responsibility, competence, and reciprocity. The Care Centered Framework by van Wynsberghe deviates from Tronto's model regarding the value of reciprocity. It highlights the bidirectional relationship between caregiver and care recipient rather than emphasizing responsiveness. While the values of attentiveness, responsibility, competence are ascribed to caregivers, the value of reciprocity is attributed to care-receivers. To ease the application of the framework, van Wynsberghe proposed the Care Centered Value Sensitive Design Methodology (CCVSD) [55] and used it to inform the conceptual design of a robot prototype to highlight the problematics linked to moral agency and responsibility [56].

In a similar vein, Sorell and Draper [41] proposed an ethical framework for designing robot companions for older adults living independently based on a review of existing literature on care robots and ethical considerations in elder care. Their framework suggests six values to be considered: autonomy, safety, enablement, independence, privacy, and social connectedness. In particular, the authors prioritized autonomy above all other values because they considered autonomy the most important ethical value in the design of care robots for older adults for domestic use. The framework is evaluated through an international focus group study with older adults, informal caregivers, and formal caregivers [57]. The authors presented participants with scenarios involving potential conflicts between these values and found that participants confirmed the prioritization of autonomy, except in cases where safety was at risk [57].

While the above introduced frameworks rely only on predefined values, Umbrello and van Poel [46], inspired by CCVSD [55], developed the AI4SG methodology. This methodology proposes examining three categories of values in a specific order: (1) values to be promoted, (2) values to be respected, and (3) context-specific values to be derived. They argued that incorporating Artificial Intelligence into SARs introduces novel challenges, such as big tech companies reshaping the healthcare landscape. Therefore, a broader set of values should be considered in AI systems such as SARs. For values to be promoted, the authors referenced the UN Sustainable Development Goals [58], while for values to be respected, they highlighted ethical guidelines developed by the European Commission [59]. To illustrate their methodology, Umbrello and van Poel [46] conceptually investigated the design of a SAR for a hospital setting. In particular, they identified context-specific values through a conceptual analysis after determining the values to be promoted and respected.

While the context-specific values play a subordinate role in the AI4SG methodology, other frameworks highlight the importance of considering the given value context for designing SAR solutions. These frameworks specifically emphasizes the usage of empirical methods (e.g., surveys, interviews) to gather stakeholder input on context-relevant values [60, 61]. This approach is particularly important when it comes to resolving value tensions. Value tensions can arise at various levels, for example, between stakeholder groups, before technically implementing value-sensitive SARs [62–64].

While the majority of conceptual frameworks approach values in HRI design from an ethical perspective, Čaić et al. [65] propose a distinct framework. Drawing from service, robotics, and social cognition research, their model seeks to understand the perceived benefits or utility of SARs in service contexts. The authors argued that the perceived benefits or utility of SARs in elderly care services depend on the alignment between the robot's functionalities and the personal values of its users [28]. This evaluation is mediated by the robot's perceived warmth and competence, which are conveyed through its affective and cognitive capabilities.

## 2.2. Empirical Approaches to Value-Sensitive HRI

While numerous frameworks and design approaches exist for supporting designers in embedding values in HRI, only a few empirical studies examine values in HRI design. Similar to the conceptual approach, existing empirical works also have a strong background in ethics.

While it is paramount to investigate how certain values, such as privacy [66–68] and safety [69], are interpreted in the care context, research has also focused on understanding ethical concerns, that is, the fear of a lack of values, in aged care to inform HRI design [70–72]. Both Hung et al. [70] and Wangmo et al. [71] explored potential ethical concerns related to the use of SARs in dementia care using qualitative methods. While Wangmo et al. [71] considered the view of healthcare researchers and professionals, Hung et al. [70] also involved older adults and their families. Both studies share the findings that the stakeholders are concerned about privacy, justice, and loss of human connections, and emphasized that an ethically aligned design of SARs is needed in dementia care.

In a similar vein, Kim et al. [72] asked caregivers and medical staff to assess dignity, control, autonomy, and privacy within two ethical dilemmas centered on the use case robots. The first dilemma examined the tension between an older adult's need for fall detection monitoring in the bathroom (enabled by image-sensing) and their desire for privacy. The second dilemma involved an older adult not taking medication despite a robot's reminders. While both groups reached similar ethical decisions in the second dilemma, their value prioritization diverged from the first. Medical staff expressed concern regarding the impact of filming on the older adult's dignity and immediate respect for autonomy when the individual requested to stop filming. Conversely, caregivers perceived observation as necessary for fulfilling their care duties, even with potential infringement upon the older adult's dignity and autonomy. These findings subsequently informed the design of pose-sensing algorithms that analyze joint movements from camera images instead of storing camera data for the robot's intended services.

Beyond ethical considerations, Kim et al. [73] focused on social values of care robots, which is conceptualized as the positive impacts that contribute to the overall well-being of society. Focus group interviews with caregivers and care recipients revealed ten values, which were clustered in four value categories: labor (e.g., competent caregiving, essential care provision), health (e.g., autonomy, independent living), innovation (technological and organizational), and economics (cost-benefit considerations). These values were then ranked by experts from the care service and care robot industries through an Analytic Hierarchy Process. Notably, technological innovation and labor/health benefits were ranked highest, while autonomy and needs as well as development and management costs were ranked lower. The study concluded with the importance of considering social values alongside economic factors in SAR design.

While existing research provides a solid foundation for addressing values in HRI design, several limitations are evident. Firstly, it can be observed that conceptual frameworks dominate the current research landscape. More empirical work is needed to understand how stakeholder values manifest in real-world settings, especially with regard to potential value tensions [72]. Secondly, most of these frameworks take an ethical standpoint and provide pre-defined values [40, 46, 57]. While these values can serve as a starting point for integrating values in HRI, there is a growing trend towards broadening the view of values [60, 65, 73]. In particular, researchers increasingly emphasize the need for investigating values specific to the context [38, 39, 74]. Finally, current research primarily focuses on the perspectives of caregivers and care-receivers, neglecting the broader range of stakeholders potentially impacted by SARs. For instance, residents in nursing homes might inadvertently disclose information about relatives, compromising their privacy. Thus, a multi-stakeholder view is essential to comprehensively consider all relevant values in HRI design [75, 76]. The presented work contributes to the limited corpus of empirical studies, emphasizing context-specific values and incorporating a diverse set of stakeholders.

# 3. Method

The presented work builds on the VSD approach [29, 60]. We deliberately chose this approach because it underscores the profound connection between technology and human values, emphasizing the need to integrate values into the technology design. VSD particularly stresses the importance of considering both direct users and those indirectly affected by the technology [29]. In line with these principles, we investigated values deemed important by stakeholders in the context of a SAR designed for use within a nursing home setting.

The aim of the SAR application that we investigated is to improve the quality of life for residents by ensuring they arrive at scheduled therapy sessions on time, while also reducing the workload of nursing home staff. For this purpose, the robot is intended to perform six tasks, including receiving the assignment, locating the resident, reminding the resident about the therapy appointment, guiding the resident to the therapy room, maintaining contact with the resident throughout the journey, handing over the resident at the therapy room.

## 3.1. Participant Recruitment

A stakeholder analysis [29] was conceptually conducted for this use case. Residents, caregivers, and therapists were identified as direct stakeholders due to their immediate interaction with the companion robot. Indirect stakeholders included relatives of residents and other nursing home staff, as the companion robot's deployment could potentially have implications for their experiences. For instance, relatives might express safety concerns regarding their loved ones, and while other nursing home staff may not directly utilize the robot, its presence would shape their workplace environment. Consequently, the values held by indirect stakeholders were deemed equally important.

Participants were recruited through two distinct methods. Researchers extended invitations to residents and those present as bystanders following an informational session outlining the research project. The remaining participants were recruited via a designated contact person at the elderly home. Table 1 provides an overview of the recruited participants, categorized by their type of involvement and providing demographic information.

*Table 1 Overview of Stakeholder Groups*

| Stakeholder Group | Stakeholder Type | Number of Participant | Gender ratio (f = female, m = male) |
|---|---|---|---|
| Resident | Direct | 5 | 3 f 2 m |
| Care givers | Direct | 8 | 3 f 5 m |
| Therapist | Direct | 5 | 3 f 2 m |
| Relatives | Indirect | 3 | 1 f 2 m |
| Bystander | Indirect | 2 | 2 f |

## 3.2. Data Collection Process

This study employed two qualitative methods for data collection. Value-oriented semi-structured interviews [29] were chosen as a method because it allows interviewees to provide in-depth explanations of their values, beliefs, and motivations. Moreover, the semi-structured format gives researchers the flexibility to follow up on interesting or unexpected points raised by the interviewee. This can lead to the discovery of values or concerns that weren't initially anticipated by the researcher, offering a richer dataset.

To elicit stakeholders' perspectives on the use case, a value-oriented interview guideline was developed to structure the interview around the six tasks of the SAR application. Overall, the interviews were structured in two phases. First, the interview focused on deriving stakeholder values by asking open questions, such as what participants deem important in the particular context (e.g., SAR reminds resident of therapy appointment), reflecting both current practices (without SAR) and future scenarios (with SAR) for each task. Second, participants were presented with a curated list of 14 values (see Table 2). Each of these values was printed on a card with a clear definition. Participants were asked to select three values they considered most important for each specific task and explain their reasoning.

The list was compiled from previous works like the moral elements of the Care Centered Framework [40], the human values with ethical import as suggested by VSD [29], the values proposed in the International Charter for Human Values in Healthcare [77], and finally, the four values to be respected in AI systems according to Umbrello and van Poel [46]. We carefully considered the relevance of each value to the presented use case when consolidating the list. Similar values were aggregated (e.g., right to equality and freedom from bias combined as fairness) while those irrelevant to the use case (e.g., environmental sustainability) were omitted.

*Table 2 Values Derived from the Literature and Presented in Interviews after Open Questions*

| Value | Definition |
|---|---|
| Autonomy | Refers to people's ability to make independent decisions, plan and act in such a way that they can achieve their own goals. |
| Calmness | Refers to a relaxed and calm state or impression. |
| Empathy | Refers to sympathy for the suffering, distress or similar of others. |
| Competence | Refers to the ability to take action to meet the needs of others. |
| Courtesy | Refers to friendly and courteous treatment of others. |
| Curiosity | Refers to the desire or urge to learn about things that are unknown. |
| Fairness | Refers to decent behavior; fair, honest attitude towards others. |
| Helpfulness | Refers to the willingness to help others, to be helpful. |

| Privacy | Refers to the legally protected, personal living environment. |
|---|---|
| Respect | Refers to respect and appreciation for someone/something. |
| Responsibility | Refers to the obligation to ensure that everything goes as well as possible, that what is necessary, and right is done and that no harm is done. |
| Safety | Refers to the state in which one is protected from danger. |
| Trust | Refers to being firmly convinced of the reliability, dependability of a person or thing. |
| Wellbeing | Refers to good physical and mental health. |

The interview guideline underwent pilot testing for refinement. All interviews took place at the elderly home, each with a duration of 30-45 minutes. With participant consent, all interviews were audio-recorded for subsequent analysis.

In addition to the interviews, two focus groups were conducted: one with the care staff (N = 5), therapist (N = 3) and bystander (N = 1), and one with the residents (N = 3) accompanied by care staff (N = 1). We chose this method because it allows us to gain insights into different perspectives through discussion among participants [78]. In the focus group with the care staff and therapist, the anticipated benefits and challenges associated with the robot-assisted accompaniment were explored. In the focus group with the residents, various potential interactions with the robot in the use case were demonstrated using an actual robot, and the residents' preferences and reactions were discussed. Additionally, the residents were also invited to share their expectations regarding the use case. Both workshops were held on site in a dedicated room at the nursing home and lasted 90 minutes. With participant consent, the focus groups were recorded on video and audio.

### 3.3. Data Analysis

All interviews and workshops were transcribed and coded in the qualitative data analysis software MAXQDA 2022. An inductive approach to the data analysis was employed. To begin, one researcher thoroughly read the transcripts for familiarization. Open coding was used, defining individual words, word groups, and sentences as the unit of analysis [80]. Similar to [79], initial codes regarding participants' views of a robot-assisted accompaniment were generated and iteratively refined through a process of merging similar codes. For example, the initial code "Getting to the therapy room without accidents" was merged with the initial code "Ensuring safety by providing guidance to the destination", resulting in the code "Protecting resident from danger on the way to the therapy room". Overall, a total of 315 initial codes were generated and merged into a final set of 104 codes. Next, similar codes were grouped into themes [79]. For instance, "Personalized walking behavior" and "Protecting resident from danger on the way to the therapy room" are grouped under the theme "Robot in Motion".

To identify stakeholder values, we first considered values explicitly selected by interview participants and directly assigned them to the corresponding codes. Additionally, guided by the consolidated list, we extracted values from participants' responses to open-ended questions. If the coded data did not match existing values, new values were introduced. For example, when interviewees reported that the robot should refer to the resident's biography during small talk linking this to the values of "curiosity" (from the curated list), we also assigned "individuality" (a new value) to this coded data to reflect the stakeholder's need for the robot to recognize and respond to each resident's unique life, emphasizing personalized interactions. To ensure the reliability of the extracted values, a separate team member evaluated the value coding, which included both new and literature-based values. This team member independently evaluated the 315 initial codes, assigning values according to the list and suggesting any additional values as needed. Any coding disagreements were resolved through discussion.

## 4. Results

This section presents the values identified from the interviews and workshops. We start by presenting the relevant values, followed by an illustration of the distribution of values mentioned by the participants. Then we will elaborate on the values in detail.

In total, twenty-two values are identified to be relevant in the usage context. All values stemmed from literature and presented to the interviewees were at least selected once. In addition, we identified eight further values from participants' responses (see Table 3).

*Table 3 Stakeholder Values Relevant for the Usage Context*

| Value | Definition | Source |
|---|---|---|
| Autonomy | Refers to people's ability to make independent decisions, plan and act in such a way that they can achieve their own goals. | [29, 46, 77] |
| Calmness | Refers to a relaxed and calm state or impression. | [29] |
| Empathy | Refers to sympathy for the suffering, distress or similar of others. | [40, 77] |
| Competence | Refers to the ability to take action to meet the needs of others. | [40, 46, 77] |
| Courtesy | Refers to friendly and courteous treatment of others. | [29] |
| Curiosity | Refers to the desire or urge to learn about things that are unknown. | [77] |
| Fairness | Refers to decent behavior; fair, honest attitude towards others. | [29, 46, 77] |
| Helpfulness | Refers to the willingness to help others, to be helpful. | [77] |

| Value | Definition | Source |
|---|---|---|
| Privacy | Refers to the legally protected, personal living environment. | [29, 77] |
| Respect | Refers to respect and appreciation for someone/something. | [29, 77] |
| Responsibility | Refers to the obligation to ensure that everything goes as well as possible, that what is necessary, and right is done and that no harm is done. | [29, 40, 77] |
| Safety | Refers to the state in which one is protected from danger. | [46, 77] |
| Trust | Refers to being firmly convinced of the reliability, dependability of a person or thing. | [29, 77] |
| Wellbeing | Refers to good physical and mental health. | [29, 77] |
| Care | Refers to emotional, physical, and psychological support and consideration for the well-being of others | Empirically identified |
| Collaboration | Refers to cooperation to achieve common goals | Empirically identified |
| Efficiency | Refers to accomplishing tasks in high quality and in the shortest possible time with minimal resources | Empirically identified |
| Enjoyment | Refers to offering pleasure and joy | Empirically identified |
| Individuality | Refers to ensuring individual needs and preferences | Empirically identified |
| Perceived Control | Refers to being able to determine or influence important events or situations | Empirically identified |
| Humor | Refers to the use of humorous expressions | Added by BystandersP1 |
| Mindfulness | Refers to following events with increased attention | Added by CaregiversP17 |

Figure 1 illustrates the distribution of values mentioned by participants. *Competence* emerges as the most prominent value, followed by *individuality*, *trust*, and *safety*. *Privacy* and *collaboration* received similar levels of mention. From *responsibility* onward, a gradual decline in the frequency of values is noticeable. However, starting at *autonomy*, the decline becomes sharper, with values such as *humor* appearing the least frequently.

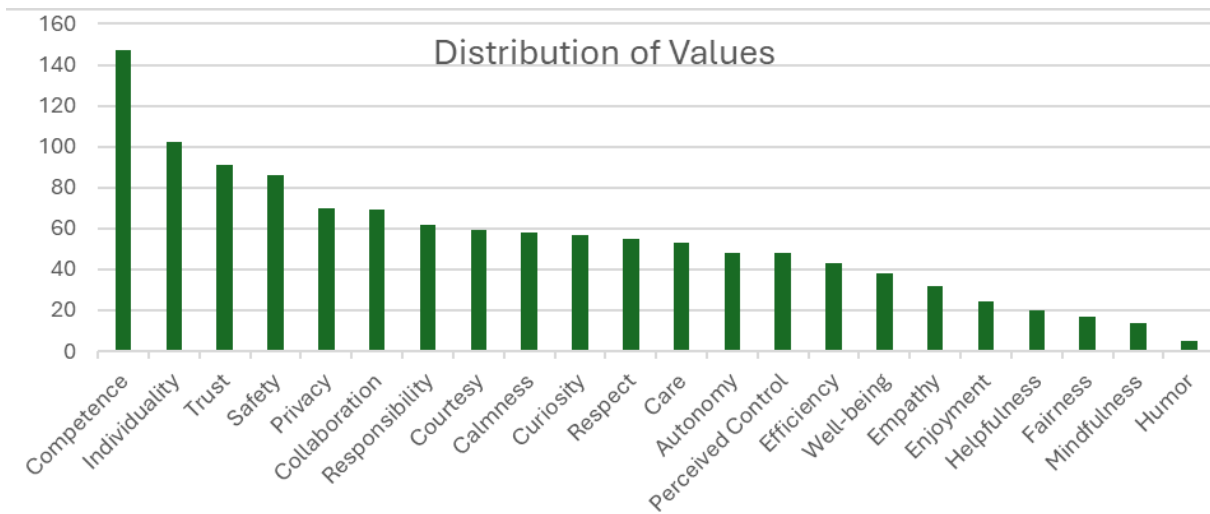

*Fig. 1 Distribution of Values based on Their Mentions*

Following the structuring approach outlined by Köhler et al. [81], we organized the results into five core values to enable a comprehensive understanding of how different values are intertwined with SAR use: safety, autonomy, privacy, care, and trust. Each core value reflects the way stakeholders perceive values within the context of SAR usage in the nursing home environment. In line with Draper and Sorell [57], the first three core values—*safety*, *autonomy*, and *privacy*—cover critical aspects such as ensuring physical and psychological safety, respecting residents' autonomy in decision-making, and safeguarding their informational, physical, and social privacy. Following Köhler et al. [81], we further grouped values that have a synergistic effect under the core value of *care*. This core value encompasses *individuality*, *courtesy*, *calmness*, *respect*, *empathy*, *helpfulness*, and *fairness*, all of which are essential to providing effective emotional, physical, and psychological support [82, 83] and contribute directly to promoting the well-being of residents, making the value of *well-being* a natural extension of this core value. Given that *care* encompasses practical actions designed to enhance well-being, *efficiency* can be regarded as a distinctive quality of care actions. Finally, the core value of *trust* comprises of values such as *competence*, *cooperation*, *responsibility*, and *perceived control*, as they frequently co-occur with trust as indicated in our data and are an integral part for building trust [84, 85]. Since the values of *mindfulness*, *enjoyment* and *humor* received little attention in the data, they are not included in the detailed elaboration in the presented work, allowing a focused discussion on the most significant aspects of the findings.

### 4.1. Safety

An important value that emerged from the data is the *safety* of residents, as put by one caregiver: *"[t]he safety of the residents is the top priority for the care team to ensure their well-being at all times" (CaregiversP8)*. The concept of *safety* in HRI can be divided into two categories: physical and psychological safety. *Physical safety* is defined as the prevention of harm through physical contact, whereas *psychological safety* refers to the avoidance of psychological harm, such as stress and discomfort, on human beings. [86]. When stakeholders elaborate on *safety*, *trust* is the most associated concept. In particular, stakeholders recognized the importance of addressing the two categories of safety in order to foster trust in the robot.

With regard to the *physical safety* of the robot, the analysis indicates that the primary concern is the task of guiding the accompanying person to the therapy room. In particular, all stakeholder groups are severely concerned that residents may be at risk of falling on their way to the therapy room for various reasons. These reasons include a sudden change in physical fitness, stumbling over objects on the floor, and colliding with objects. To mitigate this potential hazard in the robot-assisted scenario, two kinds of measures are needed from the stakeholder perspective.

Firstly, in the current practice, caregivers and therapists provide physical support, allowing residents to lean on for stability during walks. In analogue, a robust robot design capable of providing the same support for residents is desired, as exemplified by TherapistsP13: "*[it is important] that you can lean on [the robot], for example. It happens all the time that someone stops and says: "I feel dizzy". Then [the robot] has to be able to support them*". Secondly, stakeholders viewed several *competences* necessary to ensure safety of residents, which can include accompanying the resident with clear guiding instructions of the path, personalizing the robot's motion according to the resident's speed and raising emergency alerts in case of an incident. Such incidents include a resident with sudden health-related needs or the robot experiencing unexpected issues and therefore requiring the caregiver to take over. These proposed measures underscore the necessity for the robot to recognize residents' *individual conditions*, while equipping with the capacity for effective *collaboration* with care staff in order to ensure the safety of residents in the task of guiding the accompanying person to the therapy room.

*Physical safety* is also evident in the task of locating the accompanying person. When searching for a resident, a robot might encounter other residents walking in the hallway. While the residents find it acceptable to be overtaken by the robot, the robot must ensure that it overtakes them safely and does not push them aside. In particular, the residents express the desire that the robot should overtake from the resident's left side, adjust its speed, and give a signal while overtaking. Similarly, after successfully locating the resident, the robot should approach the resident in a calm manner. These measures would make the residents aware of the robot in time to avoid an unpleasant surprise. Due to their vulnerability, any unexpected fright could jeopardize the safety of the residents.

With regard to *psychological safety*, caregivers, therapists and relatives voiced the need to provide a sense of security to the residents to avoid discomfort. This can be done by, for example, assuring the accompaniment to the therapy room and conveying self-confidence that the robot can successfully accompany the resident to the therapy session, as explained by RelativesP12: *"[The robot] comes into the room and says: 'You have an appointment then and then, do you need anything else?' … Then say again: 'I'm coming with you now and will accompany you' to ensure a sense of security. The robot should convey the feeling that it knows where it is going."*

## 4.2. Autonomy

All stakeholders associated *autonomy* with resident's ability to make own decisions and act accordingly [29], and stressed that the autonomy of the residents should be respected. This is particularly evident when the resident refuses to attend the therapy session when reminded by the robot, as emphasized by one of the exemplary statements: "*[t]]he person should be allowed to decide autonomously. The robot must be able to motivate [the resident for therapy], but there are limits somewhere, ultimately the human decides. It should motivate but not decide.*" (RelativesP12)

Indeed, all stakeholder groups except the resident group emphasized that the robot should motivate the resident to participate in the therapy session before finally accepting the resident's refusal. This insistence on motivation is strongly motivated by the value of *care*. From the perspective of caregivers, therapists, relatives, and bystanders, ensuring residents attend the therapy session represents a practical action *[40]* to promote their well-being. Thus, a tension between autonomy and care emerges. However, this tension resolves itself as all stakeholders agree that residents' autonomy should be prioritized over care.

Moreover, residents emphasized the importance of having a voice in determining which robot's services are to be used, emphasizing the consideration of residents' *individuality*. This includes deciding on the extent of its involvement in their care routine, such as whether it should solely serve as a reminder for therapy sessions, or additionally accompany them to and from the therapy room. In particular, residents provided different reasons for returning to the room alone. While ResidentsP05 is reluctant to "*tell the robot where … to go*", ResidentsP19 does not want to "*burden [the robot] too much*".

However, the caregiver group raises concerns about the safety of residents who decide to return to the room on their own, as stated by CaregiversP8: "*[d]epending on the situation, the resident should not be allowed to return to the room alone. Some residents cannot assess the danger and accidents can therefore happen. Care staff cannot find the resident afterwards. It has to do with the safety and autonomy of the*

*resident. In this case, safety comes first.*" In this regard, a value tension between residents' autonomy and safety arises. Rather than simply trading off between the two, designers are encouraged to seek solutions that preserve residents' sense of independence while ensuring their safety, thus preventing feelings of disempowerment or loss of autonomy.

### 4.3. Privacy

All stakeholder groups express *privacy* concerns that can be categorized into three dimensions that affect different tasks. The first dimension is informational privacy, which refers to the right of individuals to control the disclosure of their personal information to third parties, including the method, timing, and scope of disclosure [87]. This dimension mainly concerns the verbal communication of the robot.

While all stakeholders stress that the conversation should be personalized, involving the use of the resident's personal information, to create an entertaining and pleasant atmosphere on the way to the therapy session, they want to prevent the robot from unintentionally disclosing the resident's sensitive personal information. Because of this risk, one therapist would even deny the robot's access to health information. While residents seem willing to allow the robot to access information about their biography (such as hobbies and previous jobs) for tailored small talk, including information about the resident's family in a personalized conversation is considered problematic for some caregivers, as shown in this exemplary statement from CaregiversP10: "*Personal topics such as family or health should not be discussed in the hallway. It should be a public setting in the hallway where only trivial topics are talked about*". Further, it can be observed that all members of the relative stakeholder group dislike the idea of giving the robot access to the resident's family information. RelativesP14 explained the reason: "*[The robot] shouldn't get too personal. It's better not to talk about the relatives. That could be misinterpreted under certain circumstances. It should be trivial, where no one can be offended or hurt.*"

The observed disinclination to share family information could be explained by mistrust in the robot's ability to handle this conversation topic properly and the perceived loss of control over data sharing, which violates relatives' need for informational privacy.

Interestingly, both a bystander and a relative suggest that the robot should spontaneously generate topics and tailor the conversation rather than directly referencing the resident's biography, as this approach would help avoid potentially awkward situations and ensure a privacy-conform interaction for the resident, as elaborated by BystandersP1: "*the information should not be fed in advance based on data in order to avoid irritation (e.g. [the robot] knows where I live without me having told*

*it), but should be learned through conversation. I am prepared to disclose what I tell. It's different when the information is fed in by a third party."*

Regardless of how the robot personalized the conversation, the robot should ensure confidentiality of the conversations with the residents and not disclose information to third party. A resident ascribes this aspect essential for building *trust* with the robot: *"[Pointing to the card labeled with Trust] I am convinced that if I tell the robot something, it will remain confidential."* (ResidentsP5)

Further, few stakeholders in the caregiver, therapist and relative allow the robot to use residents' personal information (e.g., individual progress in the therapy) to motivate them for the therapy session. On the other hand, one caregiver warned for misusing health information for motivation.

These multifaceted views suggest a value tension between privacy and individuality in the form of personalized conversations and motivations. While personalization could enhance the quality of communication and care, stakeholders prioritized privacy over individuality due to the perceived risks of data breaches and a lack of trust in the robot's capabilities.

The second dimension of privacy prevalent in our findings concerns physical privacy, which is defined as "the degree to which one is physically accessible to others. This has to do with concepts of personal space and territoriality" [87]. This type of privacy affects access to the resident's room. While residents are generally willing to allow the robot access to their room, they seem to have different preferences regarding the robot's respect for their physical privacy as illustrated in the statements below. While for some residents it is sufficient for the robot to knock on the door and enter the room, another resident requires the robot to ask for permission, indicating a need for *controlling* his personal space.

*"[The robot] can search for the resident in room … Knocking before entering the room is important. Briefly explain what it is about and why the robot is there. The robot should come in so that the resident can see who is there." (ResidentsP3)*

*"The robot should not simply enter the room without permission" (ResidentsP5)*

Finally, disregarding the resident's autonomy violates the dimension of social privacy, which refers to an individual's ability to regulate his or her social interactions and control the contacts he or she has with others. The resident's control over his or her interactions with others may be compromised if the resident cannot enforce his or her decision to return alone after therapy. Similarly, another resident states that she does not want to interact with the robot on a daily basis (ResidentsP2), which may indicate a need for personal space and solitude.

## 4.4. Care

We use the value of care, referred as the combination of a person's emotional concern (caring about) and practical actions (caring for) aimed at promoting the *well-being* of others [40], to describe the various aspects that relate to how residents want to be cared for and what caregivers and therapists consider to be good care practice.

A central aspect in the value of *care* can be reflected in social intercourses between the residents and the robot. Residents found it crucial that the robot shows consideration for them, such as by inquiring about their well-being, assessing and appropriately responding to their mood. This implies that the robot should demonstrate an understanding of and sensitivity to the feelings and experiences of residents and being capable of reacting to resident's feelings. This underscores the residents' need for empathic and individualized emotional support.

*Empathy* [88] and *individuality* [81] also play an important role in motivating a resident for therapy. Caregivers and therapists emphasize that residents respond differently to different motivational strategies, and through caregiving experience, they have implicitly learned which strategy works effectively for which resident in which situation. The robot should, considering the resident's privacy, apply personalized motivation tactic and use the resident's social cues as feedback to adjust the persuasion attempt according to the situation. In this course, empathy is especially needed to accurately interpret the resident's willingness to attend the therapy session, as elaborated by TherapistsP18:

*"We always try to motivate the residents at least twice. But that's where the human element comes into play. When are the residents weighing things up and when do you have no chance? Then it might be worth saying two more sentences, depending on the situation. That's not easy. It would be sensational if [the robot] could do that, but I wonder whether it's right to recognize the nuances."*

To ensure individualized care, *curiosity* [89] is an often mentioned value when it comes to maintaining contact with the resident on the way to the therapy room. Both residents and caregivers suggested that the robot should show *curiosity* about residents' interests and adapt the conversation accordingly, while maintaining a joyful and entertaining atmosphere.

Another key aspect of care is ensuring *calmness* [29] for residents. As indicated by all stakeholder groups, calmness is particularly important due to the resident's vulnerability [90]. The value of calmness can be reflected in a variety of ways. Most residents highlighted, for example, the need of the robot being able to recognize their desire to talk and adjust its communication accordingly. ResidentsP4 gave an example: *"The robot may or may not have something to tell and should find out whether you are in the mood for a conversation and how to deal with [the situation]."* Other stakeholder groups added that the robot should give the residents enough time to do their things and interact with them in a calm and confident manner. This prevents residents from becoming overwhelmed and stressed.

*"The robot should respond slowly and directly and not react frantically, as this would overwhelm the residents. The robot should focus solely on one activity." (Caregiver, P7)*

In addition, *respect* and *courtesy* are considered essential in interacting with the robot. Residents also associate respect with recognizing their rights and allowing them to assert those rights, e.g., respecting them when they make independent decisions. In contrast, both RelativesP11 and RelativesP14 from the relatives group expressed a different perspective with regards to respect, as stated by RelativesP14: "[the robot] should also demonstrate some authority, but it must be characterized by politeness". Without this, the relatives suggested, residents may not follow the robot's guidance. *Fairness* also plays a role in the social interactions with the robot, as the residents demand that the robot treats them and other people in the nursing home decently, as stated by ResidentsP3: "The robot should treat the residents in the way that it would like to be treated … Decency in the sense of using [a formal form of interaction] when communicating".

In addition, one member of the resident group believes that it is fair to other residents if the robot does not wait until the end of the therapy and instead takes care of others who need help. In a similar vein, some stakeholders stress the importance of being *helpful* when interacting with residents. This includes helping to make residents aware of how to prepare for therapy (e.g., wearing sturdy shoes to walk to the therapy room).

Finally, while the aforementioned values mostly relate to the "caring about" aspect of care, one value that was only mentioned by caregivers and therapists in relation to "caring for" is *efficiency*. In particular, these two groups stressed that the way the robot receives the task assignment should be as efficient as possible, for example, by automatically accessing the appointments calendar. In this respect, the stakeholders' conceptualization of efficiency corresponds to the concept of efficiency in usability, which aims at minimizing task completion time [91].

### 4.5. Trust

When we curated the value list for the interviews, trust is conceptualized as a conviction that a person or thing is reliable and dependable [29, 92]. The analysis of the empirical data shows that the stakeholders shared the same understanding. Given the vulnerability inherent in elderly care, all stakeholder groups emphasize the robot must function as expected, as pointed out by a therapist: "*Trust is the prerequisite to work with [the robot]*" (TherapistsP13).

To establish trust, it is essential to execute the companion assignment reliably, for example, finding the right resident and bringing the person to the right location. Two values that frequently emerge together in this context are *competence* and *responsibility*. In our study, the value of *responsibility* is conceptualized as the obligation to ensure that all actions are carried out in the most optimal manner, that all necessary and appropriate measures are taken, and that no harm is done [93]. *Competence*, on the other hand, refers to the ability to take action to meet the needs of others [40]. The co-occurrence of these values might be explained by the fact that stakeholders require the robot to possess certain capabilities to fulfill its intended task in a consistent and satisfactory quality. Only when the robot demonstrates both competence and responsibility, trust can be established.

In this realm, the analysis suggests that the value of *collaboration* and *perceived control* can shape the expectations towards trust, competence and responsibility. Caregivers and therapists view *collaborative activities* with the robot as trust-building and essential aspects of its competence and responsibility. In this regard, caregivers expressed a desire for the robot to notify them when the status of a task is changed, for instance, whether the resident is found or not, and when the robot detects that the resident is not feeling well. Therapists echoed this desire, particularly for notifications if a resident refuses therapy attendance. This need to stay in the loop reveals the underlying value of *perceived control* [94] over irregular events, which caregivers and therapists desire to influence. Notification is an effective method for enabling them to take appropriate actions. Another example of collaborative activities that is highly appreciated by both care staff and therapists is the desire for the robot to provide reports on residents' conditions when handing over and picking up a resident, as put by CaregiversP06:

"*It is important that the report between caregivers and therapists can take place indirectly via the robot and that everyone can be informed. At handover and pick-up, the robot must record how the therapy went and pass this on to the nursing staff*".

Through the robot both care staff and therapists hope that the information exchange can happen timelier and more regularly.

Moreover, caregivers emphasized a two-way collaboration, not just the robot assisting them. They see the robot as a team member they can support, for instance, by helping the resident get dressed if the robot requires assistance. These outlined, desired functionalities highlight the importance of both *collaboration* and *perceived control* in case of deviations from routine. Providing possibilities for caregivers and therapists to collaborate and intervene as necessary strengths the trust in the robot.

The analysis of the empirical data also shows that the stakeholders' conceptualization of trust goes beyond the provided definition. This is evident when one resident expressed that "*I want to be able to trust the decision made by the robot*"

(ResidentsP4), indicating the need for building trusting relationships with the robot. Due to their fragility in cognitive abilities and physical fitness, residents are dependent on the assistance of others in daily tasks or decision-making, potentially creating insecurity. This might be one reason why the residents expected that the robot shall act in their best interests. Indeed, the need to build trusting relationships is supported by other evidence from the data. For example, as noted in the privacy section, ensuring the confidentiality of conversations is seen as central to building trust, signaling that residents are willing to expose vulnerabilities to the robot. In addition, both caregivers, therapists, and relatives acknowledge that the absence of trust could lead to residents' ignorance of the robot, as put by CaregiverP8: "*One can imagine that residents are skeptical. Trust from the residents is necessary for them to follow the robot*". In this regard, the conceptualization of trust is more in line with the definition proposed by Friedman and Hendry [29], who refer to trust as expectations between individuals, and that expectations involve good intentions, vulnerability, and the possibility of betrayal.

A closer look at the values associated with trust reveals, however, that a potential value tension might arise between collaboration with care staff and resident's privacy if the robot communicates information about residents, including physical and psychological conditions, to the care staff without their consent. While caregivers generally agreed with the confidentiality of conversations between the robot and the residents as previously described, they made an exception for information about residents' physical and psychological conditions. If not properly resolved, this value tension could result in residents feeling surveilled by the care staff through the robot, which could ultimately erode their autonomy.

# 5. Discussion

In this work, we investigated values in the context of a companion robot. Our findings reveal that the use of a SAR in this context is highly complex, touching on ethical, social, and organizational aspects. In line with previous findings [95] , the robot is viewed as both a tool and a social partner, taking on various social roles. Being a tool, the robot shall facilitate the information exchange between care staff and therapist. As a social partner, the robot is expected to act as a motivator encouraging residents' participation in therapeutic activities, entertainer providing personalized and joyful conversations, companion preventing residents from harm, and teammate collaborating with care staff and therapists. Since users apply different criteria to evaluate tools (e.g., usability [96]) and social partners (e.g., warmth [97]), this dual role might represent a barrier to user acceptance, as the robot needs to meet various criteria simultaneously.

To elicit values relevant to stakeholders, we conceptually derived values from established sources and, together with the direct and indirect stakeholders, i.e., residents, caregivers, therapists, relatives and bystander, we empirically examined how these values unfold, and other values emerge with regard to the use case.

## 5.1. Empirically Specifying Values in Elderly Care Settings

Overall, our findings confirm the relevance of the value sets we identified in the literature. In addition, eight values emerge from the data, enriching our understanding of stakeholder perspectives and stressing the importance of empirical investigation. Among these new values, we identified *individuality* (referring to ensuring individual needs and preferences) as one of the important values in designing technology for caregiving contexts. Moving beyond prior research [40, 81], we also provide empirical evidence of their prioritization within a broader set of considerations for HRI design, particularly in caregiving settings. In a similar vein, our findings reinforce the role of *enjoyment* in HRI design, which has been shown to have a positive impact on elderly people's acceptance of SARs [98]. On the other hand, values such as *collaboration*, *perceived control*, and *efficiency* emerged in relation to the caregivers and are primarily concerned with enhancing organizational performance and job satisfaction [99, 100], which affect the care quality. The diverse range of values identified in our findings underscores the necessity of empirically specifying values to address the distinct concerns of different stakeholder groups. The remainder of this section will briefly discuss the five value themes presented before and outline the empirical relevance of these values.

### 5.1.1. Trust

Trust is essential for all stakeholder groups to embrace and utilize robots effectively. Without trust, residents may be reluctant to engage with the robot, and care staff and therapists might be unwilling to use the robot, thus hindering the potential benefits it offers. In our initial definition presented to the stakeholders, trust is strongly related to the robot's reliability and dependability [29]. Prior research shows that both reliability and dependability significantly contribute to building trust towards robots [101]. In our findings, trust frequently co-occurs with *competence* and *responsibility* as indicated in Section *4.5*. Taking together, stakeholders in our study translated *trust* in the expectation that the robot takes all possible measures to ensure that the intended tasks are executed in a correct manner, e.g., delivering the reminder timely. Only when the robot exhibits its reliability and dependability through being competent and responsible, it becomes trustworthy [102] and the stakeholders can build trust towards the robot. In this regard, competence being the most frequently cited value highlights its instrumental nature and the utility aspect of the robot being a tool. When it comes to demonstrating competence and responsibility, caregivers and therapists stress the importance of collaboration and a sense of perceived control, e.g., notifying them if a resident refuses therapy attendance so that they can intervene if needed. Prior works reveal that competence can foster trust [103] and robots exhibiting cooperative behavior in a game are perceived more competent, especially in a losing situation [104]. Our findings add qualitative evidence that explain the relationship between competence/responsibility and trust, as well as the role of collaboration and perceived control in seeing a SAR as a competent and responsible team member in a care setting. Furthermore, our findings show that *trust* intertwines with other values such as *safety*, and *privacy*. In particular, both respecting privacy and ensuring safety can foster trust towards SARs. Moreover, our results also show that the conceptualization of *trust* goes beyond the robot's reliable performance and also involves the expectation that the robot acts in the residents' best interests and build trusting relationships with them, e.g., by trusting the robot will the conversation confidential. Thus, our findings highlight both the various facets and the significant challenges involved in designing trustworthy HRI. In the current robot acceptance models [105, 106], the conceptualization of trust primarily concerns the reliability aspect. This might be insufficient to fully captures the versatility of trust, indicating the need for a more comprehensive approach in measuring trust.

### 5.1.2. Care

In line with previous findings [77, 81], another set of frequently cited value by all stakeholder groups centers around the cluster of *care*, with *individuality* being the second most frequently mentioned value. Personalization is the core of good care [54]; therefore, it is expected that the HRI design across the entire use case should be guided by individuality to provide optimal care for residents. While previous research leverages individuality in designing personalized care services (e.g., entertainment, medication reminder) [107, 108] and interaction modality (e.g., visual or auditory) [109], our study sheds light on several additional dimensions of individuality in HRI. First, individuality is reflected in short-term behavioral adaptation [110], such as personalizing the robot's motion according to the resident's speed, with a focus on enhancing comfort and ease of interaction. Second, individuality plays a key role in the robot's empathic responses to residents' emotional states, providing medium-term emotional support that is sensitive to the resident's context and condition. Lastly, individuality supports cognitive adaptation, reflected in the robot's ability to employ personalized motivational strategies tailored to the unique preferences and characteristics of each resident. This cognitive adaptation evolves over time through repeated interactions and emphasizes long-term goal achievement. However, with the robot's permanent availability of individualized emotional support, designers must carefully draw a line in the HRI design to prevent residents from becoming emotionally attached to the robot [14]. Furthermore, while existing research highlights older adults' preference for socially adept behavior and pleasant interactions with SARs [111, 112], our results refine these concepts by identifying a range of interpersonal values specific such as *respect* and *courtesy* as essential components of socially adept behavior that and pleasant interactions that foster positive robot-resident relationships. Moreover, the desire for embedding these values into HRI design can imply that the stakeholders see the potential of a robot being a social actor with humanlike traits.

### 5.1.3. Autonomy and Safety

Furthermore, our findings regarding the importance of *autonomy* and *safety* align with previous research [46, 57, 72, 81, 113]. This is not surprising, given its established connection to individual well-being [114–116]. This significance may be even more salient for older adults, as declining physical and cognitive abilities can exacerbate the negative impacts of losing autonomy and risking safety. Beyond existing research on safety in human-robot interactions—such as perception-based systems (e.g., fall detection) and motion planning (e.g., path optimization) [117]—our results underscore the importance of designing robots with both robust safety mechanisms and psychological safety. This includes effective communication strategies, such as reassuring users about the presence of assistance or accompaniment. One possible reason autonomy was mentioned less frequently than safety in our study is the nature of the specific use case. Autonomy is mostly brought up when residents refuse therapy sessions, while safety concerns are relevant throughout the entire use case. Nevertheless, our findings illuminate several dimensions of autonomy that stakeholders emphasized. Notably, there was a focus on direct autonomy [118], where residents maintain ultimate authority over their decisions. Moreover, our results shed light on the specific dimensions of autonomy involved in value tensions. In both the conflicts between autonomy and care and safety respectively, the issue centers around negative autonomy—the right of individuals to make decisions free from external interference, even when such interference is well-intentioned. Moreover, the tension between autonomy and safety also concerns incapacitated autonomy [118], which occurs when individuals lack the necessary cognitive or physical capabilities to make sound decisions. Recognizing the distinct forms of autonomy at play is essential for developing a nuanced approach to resolving value tensions related to autonomy.

### 5.1.4. Privacy

Regarding *privacy*, our findings echo previous work [119, 120] by highlighting significant concerns about the use of robots in care settings, especially informational privacy. While participants recognized the potential for personal information to enhance communication, they expressed apprehensions about data handling, access, and potential disclosures. This complex view on personalization and privacy reflects the so-called paradoxes of technology, where positive and negative aspects coexist in the decision-making process, necessitating careful evaluation of values and concerns [120]. From the findings, we observe that stakeholders manage this paradox by restricting access to certain information types, such as family-related data. Furthermore, differentiated approaches to data access were observed. While all stakeholders agreed that the robot should have automated access to the appointment calendar, their opinions diverged with respect to the access to personal data for personalizing conversations between the robots and the residents. In particular, a bystander and a relative advocated for the generation of topics and the tailoring of the conversation in real time, rather than for the direct referencing of the resident's biography. This illustrates a fundamental distinction in how stakeholders view different types of data: while task-oriented data such as appointments may be seen as relatively benign, personal data is more closely associated with privacy concerns and potential vulnerabilities. This divergence highlights the complex interplay between functionality, personalization, and privacy and the need for a differentiated approach to data access. Notably, our findings identify a strong advocacy by residents' relatives for prohibiting the robot's access to family information to protect their own informational privacy. This provides empirical evidence for the need of including indirect stakeholders in the HRI design process to ensure responsible HRI that respects and protects stakeholder values.

## 5.2. Identifying Under-Researched Values in Elderly Care Settings

While our work confirms established values such as safety, autonomy, and informational privacy in SAR design for elderly care, we identify additional underexamined values such as physical and social privacy, calmness, and collaboration. These values are critical in daily interactions between residents, caregivers, therapists, and SARs. This section highlights their significance and calls for broader research to ensure SARs better support resident well-being and caregiving practices.

In addition to informational privacy, our study reveals under-researched privacy dimensions [121], such as physical (accessing the resident's room without permission) and social (regulating contact with the robot) privacy. This underscores the need for a comprehensive approach to privacy considerations in HRI design, encompassing not only informational privacy but also other privacy dimensions.

Furthermore, the emphasis on calmness as a key value in interactions with residents presents a noteworthy and somewhat unexpected finding. Calmness is neither traditionally recognized as a core value in care settings nor is considered a determining factors for HRI design and user acceptance of SARs [122]. However, the stakeholders adopt this value as a desirable trait for the robot in this particular care context, stressing that the robot shall radiate a sense of calm and be patient with the residents. This appropriation underscores that values and their interpretation are indeed context dependent. In this case, stakeholders have adapted the value of calmness to reflect the unique needs and vulnerabilities of residents. In another context such as a quiz game with older adults [123], it is absolutely imaginable that calmness might not identified as important. This context-dependent nature of values implies that relying solely on pre-existing value frameworks may not fully capture the nuances and complexities of value perception, thereby highlighting the need for empirical investigations in the identifying and conceptualization of values.

The introduction of a robot can lead to a redistribution of certain tasks (e.g., reminding and motivating residents), altering the dynamics of competence and responsibility between staff and the robot [40].This raises concerns among the caregivers about the undermining of responsibility through automation [124]. In this regard, collaboration plays a pivotal role in promoting trust for enabling a successful sharing of responsibility and competence between humans and robots [40, 125, 126], and can be seen as an value that addresses the organizational aspect of the context. Despite existing research highlighting the importance of robot-caregiver collaboration [127], this value is often overlooked in the current design landscape of SARs [124].

### 5.3. Resolving Value Tensions

Our findings reveal several value tensions. The first emerges between negative autonomy and care, particularly in the context of motivating residents to attend therapy. Residents often prefer to maintain autonomy by deciding freely whether or not to attend therapy, even when it is recommended for their physical well-being. On the other hand, care staff, therapists, and other stakeholders prioritize the residents' health and thus expect the SAR to encourage participation in therapy sessions, possibly by explaining the benefits of therapy. This creates a conflict between promoting health (care) and respecting autonomy (especially negative autonomy). SARs can facilitate a resolution by promoting positive autonomy [118], defined as the empowerment of individuals to make informed decisions with the support of others. By providing information about the benefits of therapy, SARs help residents make more informed choices without imposing a directive. However, residents may still choose to refuse therapy, and in these cases, the consensus among stakeholders is to respect residents' direct autonomy. Yet, care staff and therapists also require that SARs notify them if a resident refuses therapy, ensuring they can intervene when necessary. This compromise ensures that the value of care is upheld without fully overriding the resident's autonomy.

The second tension emerges between negative autonomy and safety, particularly regarding residents' desire to return independently after therapy sessions. Caregivers express concerns about the safety risks associated with unsupervised returns, especially for residents with impaired cognitive judgment or mobility. This tension highlights two key ethical issues: how much incapacitated autonomy should be tolerated for residents with diminished capacity, and the justification for paternalism. Paternalism defined "as an infringement on the personal freedom and autonomy of a person (or class of persons) with a beneficent or protective intent" [128]. The rationale for care staff favoring paternalistic measures, such as forbidding residents to return alone, is their responsibility to ensure residents' well-being, including safety. Furthermore, the search for missing residents requires additional work for care staff, distracting them from the care of other residents. Caregiver's view on their limited resources and fair distribution of care can also be observed in [57], in which a similar value tension has been reported in domestic elderly care setting. To resolve the tension between autonomy and safety, the robot should prioritize the resident's autonomy for returning alone to prevent the resident from feeling overruled and the associated loss of control [14]. However, the robot could negotiate with the resident [57] and, after accepting the resident's decision, notify care staff and therapist so that they can intervene if necessary. This kind of collaborative act helps balance the values of autonomy and safety.

A further value tension relates to paradoxes of technology and arises between individuality through the use of personal data in communication and privacy. This is particularly evident in relatives' strong opposition to sharing family information. Given this strong opposition [62] and the potential harm from unintentional disclosure, it is crucial to challenge the perceived necessity of actively/solely using personal data for creating an enjoyable atmosphere. Indeed, such an atmosphere can be fostered through present-focused, dynamic conversations responsive to residents' needs. Therefore, to resolve this tension, the robot should be designed to steer conversations towards current events, actively listen and tailor responses accordingly. In particular, the robot should only touch on personal topics if initiated by the resident and should redirect the conversation in another direction if sensitive information is involved (e.g., family). Similarly, the perceived need for using resident's health information to create personalized motivation shall be reconsidered. Generally, designers, engineers, and stakeholders must carefully consider the minimum information required for each task [129]. When personal or health data is necessary, strict access controls should be implemented to ensure it is used solely for the intended purpose.

Finally, a value tension can arise between residents' privacy and care staff's desire for collaboration in sharing information about residents' physical and psychological conditions. If not addressed properly, this could undermine residents' sense of autonomy and trust in the robot. Although the care staff's motivation is to provide optimal care, residents' privacy should still be prioritized. Specifically, when the robot detects that a resident is not feeling well, it can ask the resident for consent to share their health information with caregivers. This approach respects residents' privacy and autonomy while allowing the robot to keep caregivers informed when necessary.

The section illustrates the challenges of the deployment of SARs in elderly care environments, particularly in navigating value tensions between autonomy, care, safety, privacy, individuality, and collaboration. Based on our analysis, we propose potential resolutions to these tensions within our context. To effectively balance these competing values, designers can use these proposed resolutions to derive targeted design requirements. These requirements should encompass functionalities, interactions, and processes. For example, functionalities might include enabling residents to choose their mode of return after therapy, while interaction requirements could focus on providing a natural dialogue that addresses residents' return preferences. Additionally, process requirements should outline protocols for obtaining consent before sharing personal information with caregivers.

## 5.4. Theoretical Implications

The presented work makes several significant contributions to the field of SAR research. First, this study moves beyond conceptual frameworks and provides empirical evidence for relevant values and potential tensions in an elderly care context. While existing research primarily examines values from the user perspective, i.e., stakeholders directly interacting with the robot, our work also draws from the perspectives of indirect stakeholders (resident's relatives and bystanders), making it unique in the current research landscape. In particular, the stakeholder group Bystanders identifies *humor* as a novel value to be considered in the HRI design and together with the stakeholder group Relatives, they propose an alternative approach to how the robot should access residents' personal data, with the objective of personalizing the conversation en route to the therapy room. Furthermore, the involvement of the stakeholder group Relatives enhances the conceptualization of respect and elucidates that their privacy is at risk due to the robot's access to residents' family information. Our work demonstrates that indirect stakeholders provide unique perspectives that lead to more robust and socially aware HRI designs, fostering SARs that serve broader public interests and prevent harm. The rich qualitative data offers a comprehensive and nuanced understanding of the values that shape stakeholders' perceptions and expectations of the use of SARs in elderly care. This allows us to not only affirm existing values derived from theoretical frameworks (e.g., autonomy, competence), but also to explore their manifestations and provide a richer conceptualization of values (e.g., trust, privacy), which have often been addressed from a single perspective. Therefore, our work fills a notable gap in existing research and offers valuable insights for HRI design.

Second, our study demonstrates that values are interconnected. For instance, our findings acknowledge that individuality is related to the value of care. Optimal care cannot be achieved without considering the individuality of the resident. However, individuality alone is insufficient, as care requires additional interpersonal values such as empathy, respect, etc. While competence and responsibility often co-occur and are viewed as necessary aspects of care [40], our results suggest that these two values are, among others, more closely related to trust. This might be explained by the fact that competence and responsibility are essential for providing good care, which, in turn, fosters trust in the robot. This also stresses the important role of context plays in stakeholders making sense of values. The interconnectedness of values does not only reflect in supporting relationships such as safety promoting trust, but also in competing one, such as safety being in value tension with autonomy. Similarly, while the presence of privacy is beneficial for building trust, it can be in conflict with collaboration. By examining multiple values and their relationship rather than focusing on one single value, we contribute to a richer understanding of the interconnectedness of values, helping designers to navigate the complexities of integrating values in HRI.

Finally, the multi-stakeholder and empirical approach enable us to identify not only additional values that complement the curated list, but also under-researched values not previously emphasized in HRI design for elderly care settings. Since all the values in curated list were selected multiple times, the validity of the aggregated values in the curated list can be assured. the emergence of six values from our data—individuality, collaboration, perceived control, efficiency, care, and enjoyment—along with two values explicitly added by stakeholders—mindfulness and humor—highlight that the values derived from the theoretical frameworks did not fully capture the breath of values relevant in the current context. Moreover, the empirical approach has allowed us to identify under-researched values such as calmness and collaboration, which emerged as significant considerations for stakeholders in this particular elderly care setting. These learnings, on the one hand, underscore the importance of context in uncovering relevant values that might be overlooked in theoretical frameworks. On the other hand, they provide evidence for the need to broaden the view of designing values in HRI.

## 5.5. Practical Implications

The findings of this study offer practical implications for the design of SARs in nursing homes and similar settings.

The identified values, while not exhaustive, provide a solid foundation for informing HRI design that aligns with stakeholder values. Concretely, designers and engineers shall equip the robot with the necessary competence, such as through design features and algorithms, that promote trust and collaboration, ensure safety and autonomy, address the diversity of privacy dimensions, and provide personalized care support with an emphasis on calmness. The method of value hierarchy [130] might be useful to derive design requirements from these values. Designers might adapt the proposed resolution of value tensions for their context. As previously discussed, resolving value tensions can involve equipping the robot with appropriate functionalities, implementing effective interaction designs, and establishing clear processes—each of which requires distinct expertise. Therefore, researchers, designers, and developers should adopt a holistic and interdisciplinary approach to address these value tensions effectively. Furthermore, these values can be used as a benchmark for evaluating existing HRI designs. By assessing the extent to which a SAR's design aligns with these values, designers can identify areas for improvement.

Methodologically, VSD [29, 60] is perceived to be useful not only for identifying and understanding values from a multi-stakeholder perspective but also uncovering values unique to the setting. Designers and developers can leverage VSD to investigate stakeholder values in their context. In particular, given the critical role of context in shaping the perception and manifestation of values. Rather than relying solely on pre-existing value sources, designers and developers should conduct empirical research within the specific context of use. During the empirical research, we strongly encourage determining stakeholders carefully [14] and including both direct and indirect stakeholders during the investigation [29]. As demonstrated, the use of SARs can significantly affect indirect stakeholders and compromise their values. Thus, it becomes imperative for designers to broaden their scope beyond direct users. Incorporating the perspectives and concerns of these stakeholders into the design process can lead to more comprehensive and human-centered SAR applications.

## 5.6. Limitations

While our study provides rich, in-depth insights into stakeholder values regarding robot-assisted accompaniment in a nursing home setting, particularly within the Swiss context, it has certain limitations that provide opportunities for future research.

One of the limitations is the scope and representativeness of the participant sample. Our study focuses on a particular context, that is, a nursing home in Switzerland. Thus, the findings may not be generalizable to different cultural or institutional contexts. Although the identified values might be universal, the value tensions and prioritization observed may not be applicable elsewhere due to contextual differences. This limitation underscores the potential for future studies to expand the research to multiple settings in the healthcare system and multiple national contexts, thereby enriching the understanding of stakeholder values across varied environments.

Consistent with the widespread use of qualitative methods in research studying values for technology design [57, 62, 81], we employed semi-structured interviews and focus groups for the data collection as they are proven to be advantageous for exploring values in an implicit manner, especially in exploring various standpoints to uncover potential value tensions. To minimize constraints imposed by semi-structured interviews and focus groups, e.g., limited depth compared to open-ended interviews, we developed a detailed interview guideline and piloted a protocol to ensure a balanced and comprehensive exploration of all relevant topics.

In summary, the methodological approach of this study has provided valuable context-specific insights into integrating stakeholder values into the design of a companion robot for nursing homes. The limitations of our study offer various opportunities for future research, so subsequent studies can further deepen our understanding and ensure the broader applicability of findings across diverse settings.

## 6. Conclusion

This study investigated relevant stakeholder values for designing HRI for a SAR deployment in a nursing home setting, which aims to enhance residents' quality of life by ensuring timely therapy session attendance, while simultaneously reducing the workload of nursing staff. Through an empirical study involving direct and indirect stakeholders, we have identified a diverse set of values relevant to the usage context. While some of the identified values aligned with previous findings (e.g., safety, autonomy, and privacy), the empirical data also suggested values that are not traditionally considered in designing SARs in the care context. Moreover, our findings reveal that these values are deeply interrelated, while tensions can be observed between values, these values can also have a promoting relationship with each other. Our research not only contributes to the growing body of knowledge on human values in HRI design but also offers practical implications for the development of SARs. Future work will focus on the operationalization of values into concrete design features and the evaluation of the perception of value-sensitive HRI. Further, the conceptualization and prioritization of values might change over time. Therefore, more longitudinal studies are needed to investigate this effect. By addressing these areas, we aim to pave the way for the development of SARs that are not only technologically advanced but also ethically sound and socially responsible.

## Declarations

### Conflict of Interest

### Funding

This work was supported by Innosuisse – Swiss Innovation Agency through grant 101.588 IP-ICT.

### Consent to Participate